\documentclass[12pt,preprint]{aastex}

\def\s{{\rm\,s}}

\def\cm{{\rm\,cm}}
\def\cmm{{\rm cm}}
\def\m{{\rm\,m}}
\def\km{{\rm\,km}}

\def\gm{{\rm\,g}}
\def\g{{\rm\,g}}

\def\yr{{\rm\,yr}}

\def\s{\,{\rm s}}
\def\ss{{\rm s}}

\def\AU{{\rm\,AU}}

\def\rad{{\rm\,rad}}

\begin{document}

\shortauthors{Chiang}
\shorttitle{Repeated Resonance Crossings}

\title{Excitation of Orbital Eccentricities by Repeated Resonance Crossings:
Requirements}

\author{E.~I.~Chiang}

\affil{Center for Integrative Planetary Sciences\\
Astronomy Department\\
University of California at Berkeley\\
Berkeley, CA~94720, USA}

\email{echiang@astron.berkeley.edu}

\begin{abstract}
Divergent migration of planets within a viscous circumstellar disk
can engender resonance crossings and dramatic excitation of
orbital eccentricities. We provide quantitative criteria
for the viability of this mechanism. For the orbits
of two bodies to diverge, a ring of viscous material
must be shepherded between them. As the ring diffuses
in radius by virtue of its intrinsic viscosity, the
two planets are wedged further apart. The ring mass
must be smaller than the planetary masses so that the
crossing of an individual resonance lasts longer than
the resonant libration period. At the same time,
the resonance crossing cannot be of such long duration
that the disk's direct influence on the bodies' eccentricities
interferes with the resonant interaction between the two planets.
This last criterion is robustly satisfied
because resonant widths are typically tiny fractions of the orbital
radius. We evaluate our criteria not only for giant planets
within gaseous protoplanetary disks, but also
for shepherd moons that bracket narrow planetary rings
in the solar system. A shepherded ring of gas orbiting at a
distance of 1 AU from a solar-type star and
having a surface density of less than $500 \g/{\rm cm}^{2}$,
a dimensionless alpha viscosity of 0.1, and a height-to-radius
aspect ratio of 0.05 can drive two Jovian-mass planets through the
2:1 and higher-order resonances so that their eccentricities
amplify to values of several tenths. Because of the requirement
that the disk mass in the vicinity of the planets be smaller than
the planet masses, divergent resonance crossings may figure
significantly into the orbital evolution of planets during the later
stages of protoplanetary disk evolution, including the debris disk phase.
\end{abstract}

\keywords{celestial mechanics --- planetary systems --- planets and satellites:
individual (Uranus, $\epsilon$ ring) --- accretion, accretion disks}

\section{INTRODUCTION}
\label{intro}

\begin{figure}
\epsscale{1}
\plotone{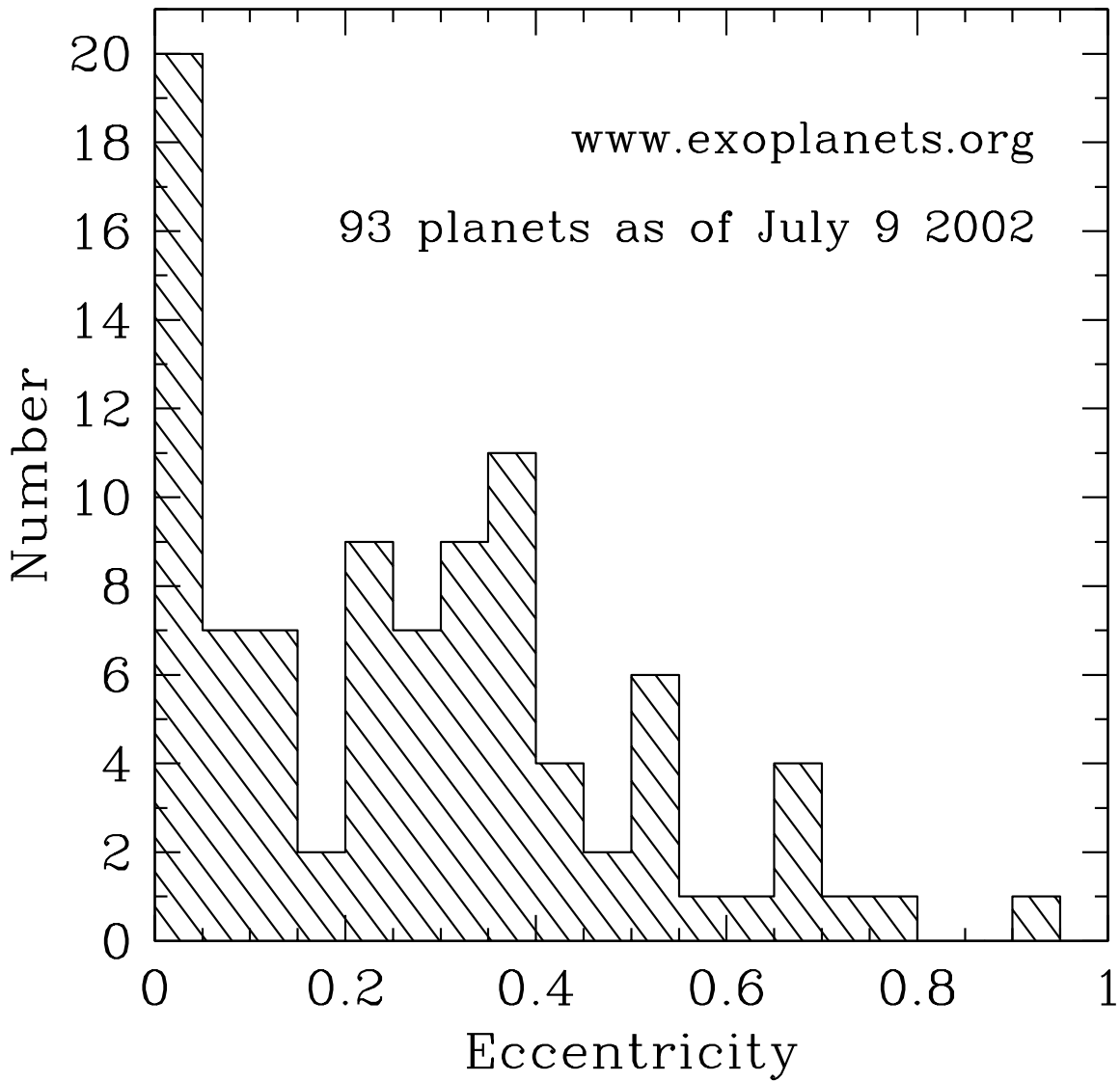}
\vspace{-1.5in}
\caption{Orbital eccentricities of 93 planets, downloaded
from the California and Carnegie Planet Search website
(http://exoplanets.org).
The spike near zero eccentricity mostly represents ``hot Jupiters''
whose orbital periods are shorter than 20 days and whose
eccentricities were likely damped by tidal interactions with
their parent stars. The remaining
distribution appears to peak near an eccentricity of 0.35.
\label{histoecc}}
\end{figure}

Orbital eccentricities of extrasolar giant planets can be
surprisingly large compared to their counterparts in the solar
system (see, e.g., the review by Marcy, Cochran, \& Mayor~2000).
Figure \ref{histoecc} displays the distribution of eccentricities
of 93 extrasolar planets, downloaded from the California
and Carnegie Planet Search website (\url{http://exoplanets.org}).
Aside from those ``hot Jupiters'' whose eccentricities
were likely damped by tidal interactions with their central stars,
most giant planets occupying stellocentric distances between
$\sim$0.2 and $\sim$2 AU
have eccentricities near 0.35, and a few have eccentricities
ranging up to 0.93.

\placefigure{fig1}

A variety of mechanisms have been introduced to excite
planetary eccentricities. These theories can be divided
into three categories: those that rely on
interactions between the planet and another point mass,
be it a star or planet; those that rely on interactions
between the planet and the circumstellar disk from
which that planet coalesced; and ``hybrid'' theories
that implicate both another body and the disk.
Kozai-resonant forcing by a binary stellar companion
(Holman, Touma, \& Tremaine 1997) and
violent encounters between two or more planets formed
within close proximity (Rasio \& Ford 1996;
Weidenschilling \& Marzari 1996; Ford, Havlickova, \& Rasio 2001;
Marzari \& Weidenschilling 2002) belong in the
first category. In the second category,
one of the most recent and inclusive studies is by Goldreich \& Sari (2002),
who demonstrate that interactions between a planet and a disk
at first-order Lindblad resonances can
excite the planet's eccentricity provided that eccentricity
exceeds a threshold value. In the third category belongs
the formation scenario proposed for the planetary system GJ 876
(Marcy et al.~2001; Lee \& Peale 2002); this scenario involves convergent
orbital migration of two planets, followed by
capture of these planets into a mean-motion resonance
and continued migration within that resonance.
This is a hybrid mechanism
because the required planetary migration is driven by an
underlying disk, while each planet's eccentricity is directly excited
by the other planet. A different
hybrid scenario may have played out for planet c in the
system Upsilon Andromedae (Butler et al.~1999; Chiang,
Tabachnik, \& Tremaine 2001). A primordial
disk may have directly excited the eccentricity
of the outermost planet, d; secular interactions
between d and c could then have siphoned off the eccentricity
of the former to grow that of the latter (Chiang \& Murray 2002).
As a bonus, this mechanism can also explain
the heretofore puzzling alignment of orbital apsides
exhibited by c and d. This process has been explored
both in the adiabatic (Chiang \& Murray 2002) and
impulsive (Malhotra 2002) limits.

Yet another hybrid mechanism has been introduced by Chiang, Fischer, \& Thommes
(2002, hereafter CFT),
who point out that if two planets migrate within a circumstellar
disk such that their orbital trajectories diverge, then their
eccentricities can increase as the planets cross
a series of mean-motion resonances. The closer the initial ratio
of orbital periods is to unity, the greater the number of resonances
that are crossed and the greater the mutual excitation of eccentricities.
Jupiter-mass planets on slowly
divergent orbits that cross the 2:1 and higher-order resonances
have their eccentricities grown from initial values $\lesssim 0.05$
to values $\sim$0.4.
Chiang \& Murray (2002) discuss
in which of the known multiple planet systems divergent resonance crossings
might have occurred.

In initially proposing the mechanism of repeated resonance crossings
via divergent migration, CFT ignored the disk's
direct effects on the planets' eccentricities. This follow-up paper
to their proposal restores such effects. We
quantitatively evaluate the conditions under which divergent migration
and the excitation of eccentricities by resonance crossings
can occur. In \S\ref{req}, we provide three such necessary
requirements and evaluate them for giant planets embedded within
gaseous circumstellar disks. In \S\ref{numex}, we test
one of these requirements by numerical integration of Jovian-mass
planets on divergent trajectories. There, we also show explicitly
which mean-motion resonances are the principal amplifiers
of orbital eccentricities. In \S\ref{prsm}, we evaluate
these conditions for the comparatively tiny shepherd moons
that confine narrow planetary rings in our own solar system;
the divergent migration of ring shepherds affords the most accessible
laboratory we have for studying repeated resonance crossings.
A summary and critical discussion of our principal findings,
including directions for future research, are given in \S\ref{conc}.

\section{REQUIREMENTS FOR PLANETS}
\label{req}

We consider the planets to be embedded within a circumstellar
disk having
surface density $\Sigma$ and kinematic
viscosity $\nu = \alpha c_s h$
at stellocentric distance $r$,
where $c_s$ is the gas sound speed, $h = c_s/\Omega$ is the vertical
gas scale height, $\Omega$ is the orbital angular frequency,
and $0 < \alpha < 1$ is a dimensionless constant.
Planetary masses, orbital semi-major axes, and orbital
eccentricities are denoted by $M$, $a$, and $e$,
respectively, with subscript 1 (2) denoting the inner (outer)
planet. The masses of the disk and of the planets
are assumed to be small compared to the mass of the star, $M_{\ast}$.
In numerically evaluating quantities,
we will take $c_s = 1 \, (r/\AU)^{-9/14} \km/\s$,
$h/r = 0.05 \, (r/\AU)^{2/7}$, $\alpha = 10^{-2}$,
$M/M_{\ast} = 10^{-3}$, and $M_{\ast} = M_{\odot}$
(see, e.g., Chiang et al.~2001; Hartmann et al.~1998).

We assume that the planets open gaps about their orbits and migrate only
by dint of disk viscosity. This mode of migration is referred
to as ``Type II'' (Ward 1997). An upper limit for the gap-opening
mass, above which planets do open gaps but below which planets
might not, is given by

\begin{equation}
\label{gapmass}
\max{M_{\rm gap}} = \frac{2c_s^3}{3\Omega G} \approx 8 \, (r/\AU)^{6/7}
M_{\oplus} \, ,
\end{equation}

\noindent where $G$ is the gravitational constant
and $M_{\oplus}$ represents an Earth mass (Lin \& Papaloizou 1993;
Rafikov 2002). A planet exceeding this mass excites strongly non-linear
density waves at principal Lindblad resonances situated near the location
of the torque cut-off, i.e., having azimuthal wavenumbers
$m \approx \Omega r/c_s$. These waves shock upon launch,
releasing the angular momentum which they carry to disk gas. We define
the gap width, $w$, to be the distance between the planet and the disk
edge which it shepherds. The gap width is estimated by setting
the viscous torque, $3\pi\nu\Sigma\Omega r^2$, equal to
the total tidal torque, $\Sigma\Omega^2r^4 (r/w)^3 (M/M_{\ast})^2$,
exerted over all principal Lindblad resonances up to wavenumber
$m_{\rm max} \approx 2r/3w$ to one side of the planet:

\begin{equation}
\label{gapwidth}
\frac{w}{r} \approx 0.2 \left( \frac{10^{-2}}{\alpha} \right)^{1/3} \left(
\frac{r/h}{20} \right)^{2/3} \left( \frac{M/M_{\ast}}{10^{-3}} \right)^{2/3}
\end{equation}

\noindent (Goldreich \& Sari 2002, hereafter GS, and references
therein). This estimate
employs the standard, linear torque formula for principal Lindblad resonances
[see, e.g., equation (10) of Goldreich \& Porco 1987].

Based on the estimate of the upper bound on
$M_{\rm gap}$ given by equation (\ref{gapmass}),
we would expect
Jupiter-mass ($M = M_J = 300 M_{\oplus}$) planets at $r \lesssim 30 \AU$
to open gaps. Uranus-mass ($M = M_U = 14 M_{\oplus}$) planets
open gaps inside $r \lesssim 1 \AU$. Equation (\ref{gapmass}) notwithstanding,
bodies having $M < M_U$ can open gaps throughout the disk, depending
on the local viscosity, temperature, and surface density; more
precise estimates of $M_{\rm gap}$ are provided by Rafikov (2002;
see his Figure 5), who employs a realistic prescription of
wave dissipation by non-linear steepening to calculate the
conditions for which Type I migration solutions transition
to Type II solutions (see also Ward 1997; Ward \& Hourigan 1989;
Goodman \& Rafikov 2001).

We identify the following three requirements for divergent
resonance crossings.

\subsection{Ring Shepherding}
\label{rs}

For the ratio of the orbital period of the outer planet
to that of the inner planet to diverge, a ring of viscous
material must be shepherded between the two planets.
As the ring diffuses in radius by virtue of its intrinsic
viscosity, the two planets are wedged further apart.

The requirement of ring shepherding supersedes the requirement
originally proposed by CFT that the viscous diffusion
time, $t_D = r^2/\nu$, increase with $r$. Even if the latter
condition were satisfied, material must exist just outside
of the inner planet to drive that planet further inwards;
otherwise, disk material just outside of the outer planet
would cause the outer planet's orbit to converge upon that
of the inner one.

A ring will fail to be shepherded if it does not span enough
wave dissipation lengths. For example, if a wave that is excited
by the inner planet near the inner ring boundary propagates across
the ring to the outer ring boundary and damps there, then disk
gas at the latter edge would be forced past the outer planet.
Similar conclusions obtain for waves excited in the ring by
the outer planet. Such ``anti-shepherding'' has been
observed in numerical simulations by Bryden et al.~(2000).

How many dissipation lengths can be accommodated by the rings
of interest here?
For the giant planetary masses under consideration,
the lengthscale over which waves dissipate, $l_d$, is not
likely to greatly exceed $h$ since the waves are already marginally non-linear
at launch. The ring width is $\Delta r = a_2 - a_1 - 2w$.
Just prior to the planets crossing a $p$:$p+q$ resonance
of order $|q|$, where $p$ and $q$ are integers and $p > p+q > 0$,
we have $a_2 = [p/(p+q)]^{2/3} a_1$. Then we require

\begin{equation}
\label{delrld}
\frac{\Delta r}{l_d} \approx 20 \left[ \left( \frac{p}{p+q} \right)^{2/3} - 1 -
0.3 \left( \frac{10^{-2}}{\alpha} \right)^{1/3} \left( \frac{r/h}{20}
\right)^{2/3} \left( \frac{M/M_{\ast}}{10^{-3}} \right)^{2/3} \right]
\frac{r/h}{20} \frac{h}{l_d} \gg 1 \, .
\end{equation}

\noindent If $p=2$ and $q=-1$, then for our choice of normalizations,
$\Delta r / l_d \approx 5$. The ring would be barely wide enough
for waves not to propagate across it, and then only for
$\alpha \gtrsim 10^{-2}$. For $\alpha < 10^{-2}$, the gaps
excavated by both planets would be so wide
that no ring could exist. The simulations by Bryden et al.~(2000;
see their section 4) employ an effective $\alpha \lesssim 10^{-2}$;
it is therefore not surprising that their rings fail to be confined.
It would be interesting to repeat their simulations with
$\alpha \gtrsim 10^{-2}$.

Our calculation is sufficiently imprecise---most notably with respect
to the dissipation length, which we have taken to be
$l_d = h$---that ring confinement may be impractical for the low-order
($|q|=1$) resonant configurations that generate the largest eccentricity
jumps. Even if $w \ll r$, then $\max (\Delta r)/h = 10$.
Our uncertainty is probably best addressed by high-fidelity numerical
simulations that can follow the evolution and dissipation
of non-linear density waves.

\subsection{Slow Crossing}
\label{sc}

For two planets to mutually excite their
eccentricities by resonant interaction,
the duration of passage through the resonance
must be longer than the resonant libration period (see, e.g., Dermott,
Malhotra, \& Murray 1988):

\begin{equation}
\label{slow}
\frac{\Delta a_{\rm res}}{|\dot{a}_2 - \dot{a}_1|} \gg T_l \, ,
\end{equation}

\noindent where $\Delta a_{\rm res}$ is the width of the resonance
and $T_l$ is the resonant libration period. This requirement
is tested numerically in \S\ref{numex}.

With no important loss of generality,
let us consider the migration of the inner planet through
a resonance of order $|q|$ and take the outer planet to be
the perturber on a fixed orbit. To order of magnitude,

\begin{eqnarray}
\label{tl}
T_l \sim \frac{2\pi}{\Omega_1} \, \sqrt{\frac{M_{\ast}}{M_2}} \, e_1^{-|q|/2}
\, ,\\
\label{dares}
\Delta a_{\rm res} \sim a_1 \, \sqrt{\frac{M_2}{M_{\ast}}} \, e_1^{|q|/2} \, .
\end{eqnarray}

\noindent These expressions are appropriate for $e \lesssim 0.3$
orbits in $|q| > 1$ resonances, and $0.3 \gtrsim e \gtrsim 0.1
(M_2/M_{\ast})^{1/3} \approx 0.01 (M_2/M_J)^{1/3}$ orbits in the 2:1 resonance
[Dermott et al.~1988; Murray \& Dermott 1999; see the latter's equations
(8.58) and (8.76)]. Insertion of (\ref{tl}) and (\ref{dares})
into (\ref{slow}) yields

\begin{equation}
\label{re4}
\frac{a_1}{|\dot{a}_1|} \gg \frac{2\pi}{\Omega_1} \, \frac{M_{\ast}}{M_2} \,
e_1^{-|q|}\, .
\end{equation}

Now it is usually remarked (see, e.g., Ward 1997 and GS)
that for Type
II drift, $|a/\dot{a}| \sim a^2 / \nu$, the viscous diffusion
time of the disk. But this statement cannot be true in the limit
that the planet mass greatly exceeds the ring mass.
We generalize the Type II drift velocity by setting
the viscous torque equal to the rate of change of angular momentum
of either the planet or the ring, whichever
is more massive:

\begin{equation}
\label{truetype2}
|\dot{a}_1| \sim \frac{6\pi\nu\Sigma r}{\max \, (M_1, 2\pi\Sigma r\Delta r)} \,
{}.
\end{equation}

\noindent This expression is a rough estimate of the
instantaneous drift velocity; it can be used to
estimate the timescale for the radial dimension of the ring-planet system
to expand by distances up to but not exceeding $\Delta r$.
Ignoring differences between $a_1$ and $r$,
we employ (\ref{truetype2}) to re-write our criterion (\ref{re4}) as

\begin{equation}
\label{reslow}
\max \, (\frac{M_1}{2\pi\Sigma r\Delta r},1) \gg 0.2 \, \frac{h/r}{0.05} \,
\frac{h/\Delta r}{0.2} \, \frac{10^{-2}}{e_1^{|q|}} \, \frac{\alpha}{10^{-2}}
\, \frac{10^{-3}}{M_2/M_{\ast}} \, .
\end{equation}

\noindent Thus, for our choice of normalizations, the planet mass must exceed
the ring mass for the divergent migration to be sufficiently slow.
Note in particular the dependence of our criterion (\ref{reslow}) on
the resonance order, $|q|$.

\subsection{Fast Crossing}
\label{fc}

At the same time, the traversal of the resonance cannot occur so
slowly that the disk's direct influence on the planet's eccentricity
interferes with the resonant interaction between the two planets; in other
words,

\begin{equation}
\label{fast}
\frac{\Delta a_{\rm res}}{|\dot{a}_2 - \dot{a}_1|} \ll \left( \frac{e}{\dot{e}}
\right)_{\rm disk-induced} \, .
\end{equation}

\noindent That this inequality might fail to be true
is a concern raised by Goldreich \& Sari (2002); among the
three requirements we discuss in this paper, it is
the easiest one to satisfy.

We again restrict ourselves
to analyzing the motion of the inner planet and take
the outer planet to reside on a fixed orbit. Following
GS, and ignoring differences between
$a_1$ and $r$, we write

\begin{equation}
\label{mineedot}
\min \, \left( \frac{e_1}{\dot{e}_1} \right)_{\rm disk-induced} \sim \left(
\frac{w}{r} \right)^4 \, \frac{M_{\ast}^2}{M_1 \Sigma r^2} \, \frac{1}{\Omega}
\, .
\end{equation}

\noindent This timescale is derived by relating either
the total first-order Lindblad torque or the total
first-order co-rotation
torque to the conservation of the Jacobi constant of the planet.
It is a minimum estimate because it considers either type
of torque alone. If both types of torques operate, their
effects tend to cancel and the timescale for eccentricity
change would increase by a factor of $1/0.046 \sim 22$
(GS, and references therein). Conservatively
proceeding with the minimum timescale, we combine
equations (\ref{gapwidth}), (\ref{dares}),
(\ref{truetype2}), and (\ref{mineedot})
to re-write condition (\ref{fast}) as

\begin{equation}
\label{refast}
10^{-2} \left( \frac{\alpha}{10^{-2}} \right)^{1/3} \left( \frac{h/r}{0.05}
\right)^{2/3}
\left( \frac{M_2/M_{\ast}}{10^{-3}} \right)^{1/2} \left(
\frac{M_1/M_{\ast}}{10^{-3}} \right)^{-2/3}
\frac{e_1^{|q|/2}}{0.1} \ll 1 \, ,
\end{equation}

\noindent where we have taken $M_1 \gg 2\pi\Sigma r\Delta r$, in accordance
with requirement (\ref{reslow}). Notice that there is no
explicit dependence on the ring surface density; increasing
$\Sigma$ increases the rate of migration, but it also
increases the rate of change of eccentricity by the same
amount. Because inequality (\ref{refast}) is well
satisfied, we conclude that each resonance crossing proceeds
undistorted by the direct effects of the disk on the planet's
eccentricity. However, the disk's direct effects would dominate
while the planets migrate between resonances.

\section{NUMERICAL EXPERIMENTS}
\label{numex}

We test relation (\ref{slow}) by numerical orbit integrations
of two Jupiter-mass planets on divergent trajectories.
Planet 1, of mass $M_1 = 10^{-3} M_{\odot}$,
is taken to occupy initially an orbit of semi-major
axis 1 AU and osculating eccentricity $e_1 = 0.01$ about
a 1 $M_{\odot}$ star. Planet 2, of identical mass to planet 1,
initially occupies a co-planar orbit of semi-major axis 1.5 AU
and eccentricity $e_2 = 0.01$. The periastra of the two orbits
are initially $0\degr$ apart. Planet 1 is placed at
periastron, while planet 2 is located at apastron.
Similar initial conditions were employed by CFT;
this configuration, integrated forward by $2 \times 10^6\yr$
with no differential migration imposed, does not betray
any instability; semi-major axes remain within 1\% of their
initial values, and osculating eccentricities average to
0.025 each.

We execute 31 integrations characterized by 31
differential migration timescales. For a given integration,
following CFT, we impose a drag force $\vec{F}_{\rm drag} =
-M_1\vec{v}_1/t_{\rm drag}$
on planet 1, where $M_1$ and $\vec{v}_1$ are the mass
and instantaneous velocity of planet 1, and
$t_{\rm drag}$ is the timescale over which $a_1$ decays.
No drag force is applied to planet 2.
The drag force is imposed at $t_{\rm start} = 500\yr$
and removed at $t_{\rm stop} = t_{\rm start} + 0.2 \times t_{\rm drag}$.
This prescription reduces $a_1$ from 1 AU to 0.67 AU
nearly linearly with time; thus,
$\dot{a_1} = -1.65 \AU / t_{\rm drag}$.
In each simulation, the two planets cross the
2:1 and 3:1 resonances, among numerous higher-order resonances.
Inserting the above speed into inequality (\ref{slow}), and evaluating
$\Delta a_{\rm res}$ and $T_l$ using the analytic
expressions given in Murray \& Dermott [1999, see their
equations (8.47) and (8.58)], we calculate that
$t_{\rm drag} \gg 1 \times 10^4 \yr$ for the 2:1
resonance to excite eccentricities effectively,
assuming a pre-crossing eccentricity of 0.025.
If the 2:1 resonance is effective, and eccentricities
become excited to $\sim$0.1, then we calculate that
$t_{\rm drag} \gg 3 \times 10^4 \yr$ for the 3:1 resonance
to be similarly effective. How well do these order-of-magnitude
criteria compare with numerical experiment?

\placefigure{fig2}
\begin{figure}
\epsscale{0.95}
\vspace{-0.60in}
\plotone{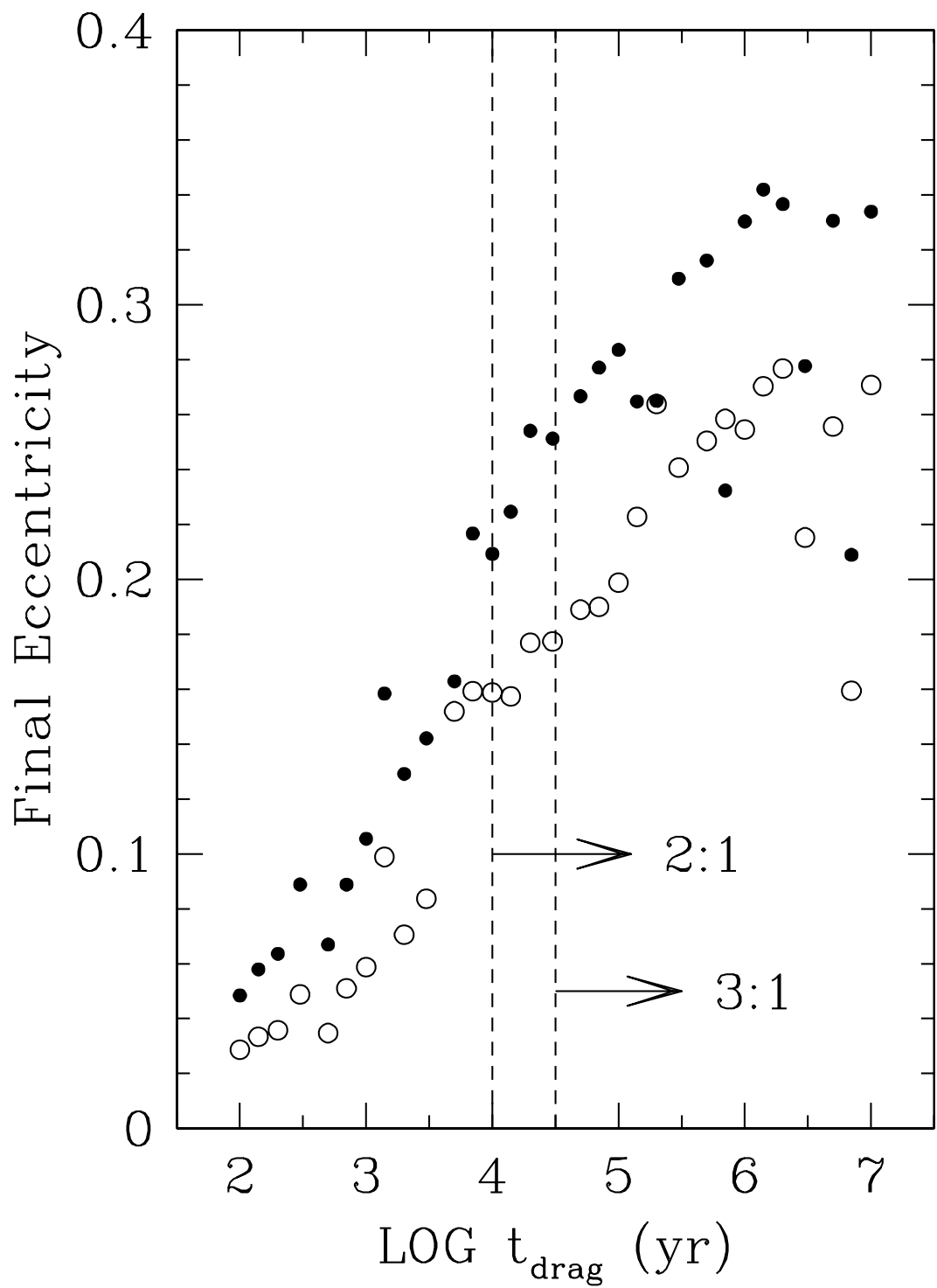}
\caption{Final eccentricities of diverging Jovian-mass planets
1 and 2, as a function of differential migration timescale,
$t_{\rm drag}$. Solid circles refer to the inner planet 1,
while open circles refer to the outer planet 2. Dotted
lines mark the minimum migration timescales required
for the 2:1 and 3:1 resonances to excite eccentricities
effectively, as estimated using relation (\ref{slow}).
\label{numex1}}
\end{figure}

Figure \ref{numex1} displays the final eccentricities
of planets 1 and 2 as a function of $t_{\rm drag}$.
The final eccentricities equal the average eccentricities
from $t = t_{stop}$ to $t = t_{stop} + 10^4 \yr$;
these final values represent averages over secular timescales.
The final eccentricities increase from $\sim$0.05
to $\sim$0.25 as $t_{\rm drag}$ increases from $10^2\yr$
to $10^5\yr$; for $t_{\rm drag} > 10^5\yr$, the final
eccentricities saturate, as we expect. We consider
Figure \ref{numex1} to largely verify relation
(\ref{slow}). Requirement (\ref{slow}) is not
as stringent as we might have guessed; even if
$t_{\rm drag} = 3 \times 10^3 \yr$, modest eccentricities
of 0.09--0.15 can still be excited.

\begin{figure}
\epsscale{1}
\vspace{-0.6in}
\plotone{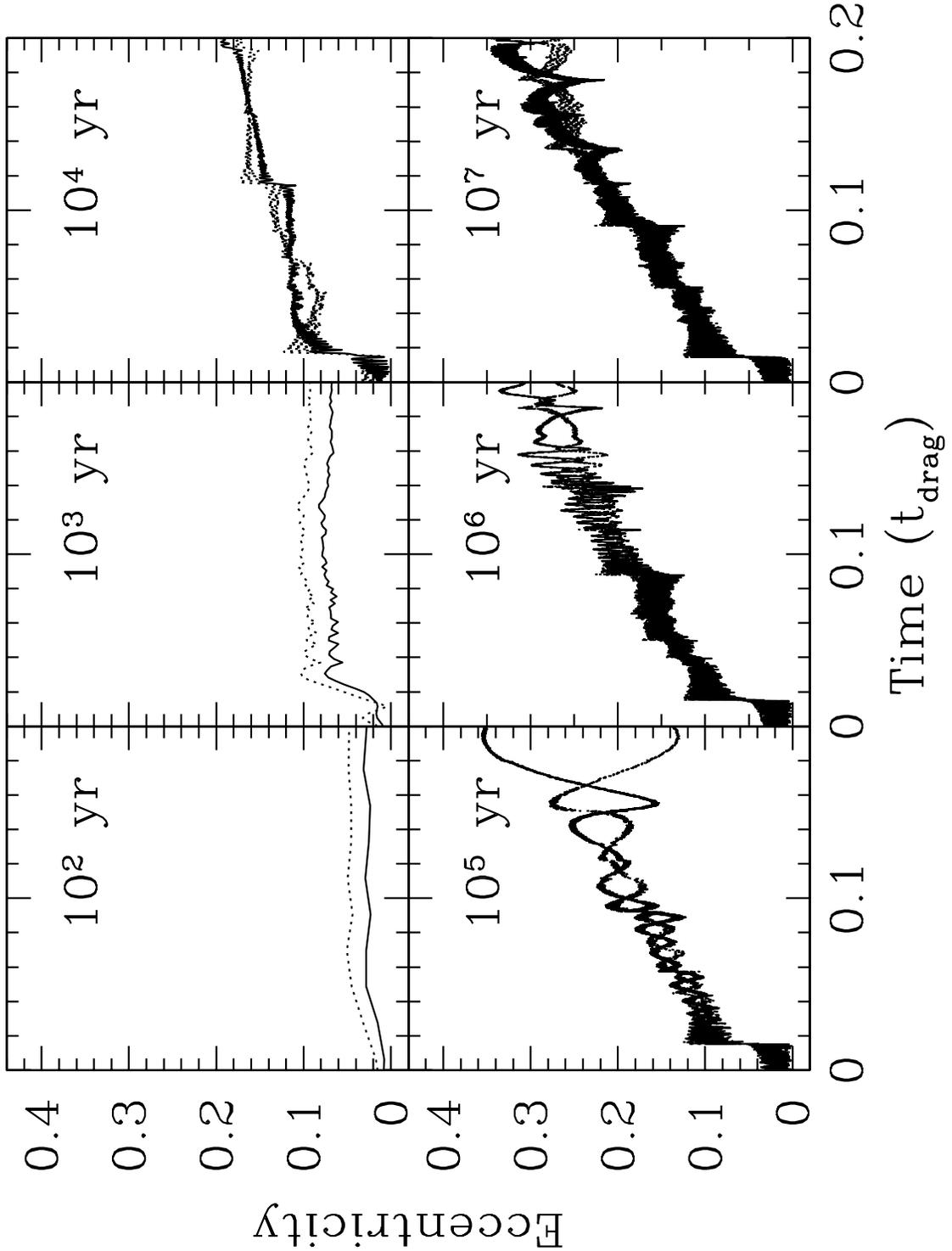}
\vspace{-0.7in}
\caption{Eccentricity evolution of diverging
Jovian-mass planets. The simulation in each panel
is characterized by a differential migration
timescale, $t_{\rm drag}$, displayed in the upper
right-hand corner. Solid lines correspond to
$e_1$, while dotted lines correspond to $e_2$.
Jumps in eccentricity
are most pronounced for $t_{\rm drag} \geq 10^4 \yr$.
These jumps are due to the crossing of numerous mean-motion resonances,
most of high-order; these resonances are identified
in Figure \ref{numex3}. Secular oscillations
in eccentricity are evident in those panels
for which $t_{\rm drag} \geq 10^5\yr$.
\label{numex2}}
\end{figure}

\begin{figure}
\epsscale{1}
\vspace{-0.6in}
\plotone{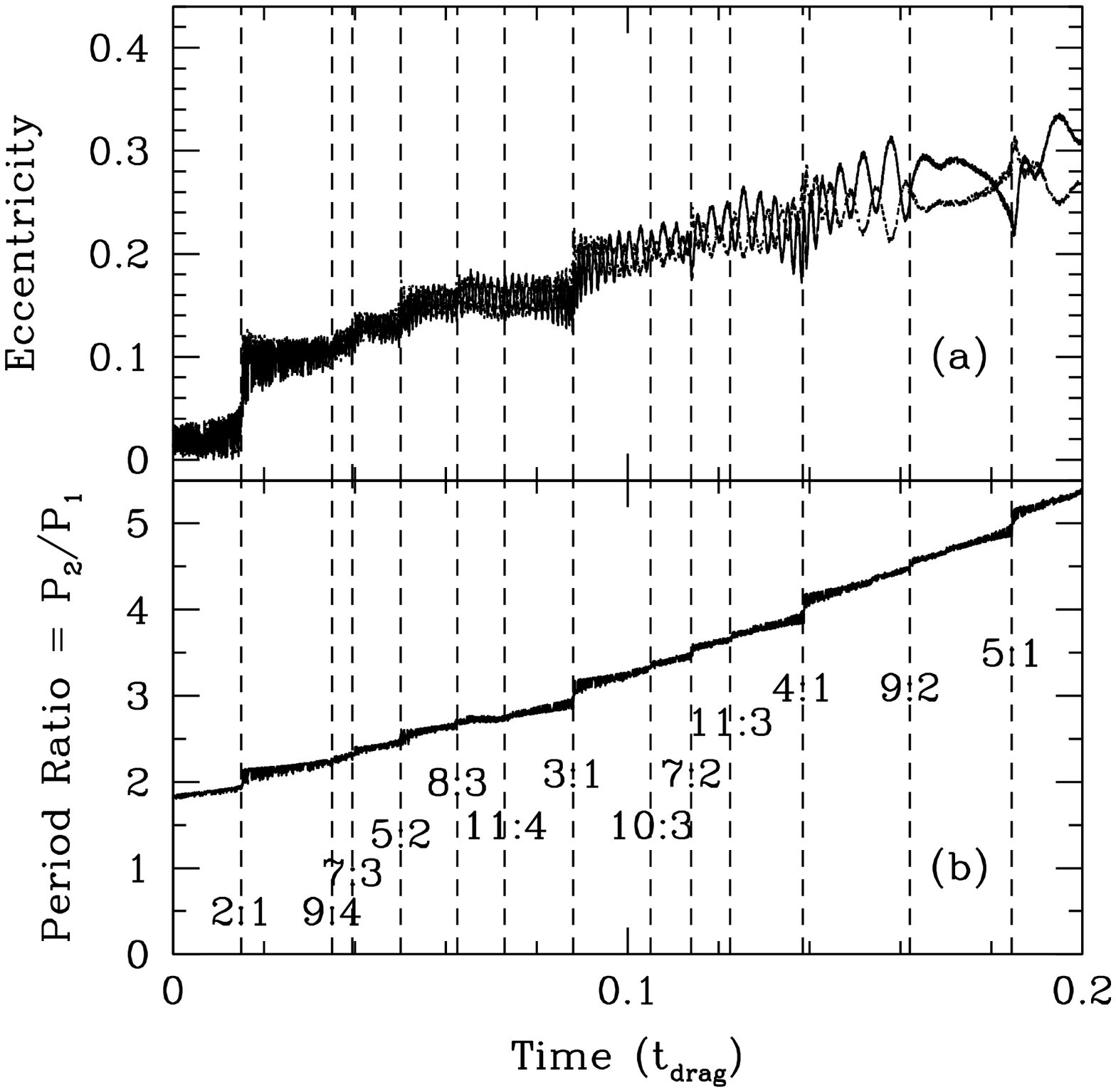}
\vspace{-0.4in}
\caption{Evolution of (a) the eccentricity
of the inner migrating planet, and (b) the ratio
of orbital periods of the outer planet
to that of the inner planet, for
$t_{\rm drag} = 10^6\yr$. Crossings of various
mean-motion resonances are indicated. Eccentricity
jumps are most marked for the 2:1,
7:3, 5:2, 8:3, 3:1, 4:1, and 5:1 resonances.
\label{numex3}}
\end{figure}

Figure \ref{numex2} displays sample time evolutions
of $e_1$ and $e_2$ for various choices of $t_{\rm drag}$.
Jumps in eccentricity that occur upon
crossing mean-motion resonances, most having $|q| > 1$,
are evident for $t_{\rm drag} > 10^4\yr$. Eccentricity
excitation is suppressed for shorter migration
timescales. For $t_{\rm drag} > 10^5\yr$,
the final eccentricities are insensitive to $t_{\rm drag}$.
Figure \ref{numex3} identifies those mean-motion resonances
that are responsible for the eccentricity jumps.

\placefigure{fig3}

\placefigure{fig4}

\section{PROSPECTS FOR PLANETARY RINGS AND SHEPHERD MOONS}
\label{prsm}

Narrow planetary rings shepherded by satellites potentially furnish
a miniature stage upon which the processes of
divergent migration and repeated resonance
crossings unfold. We take the
$\epsilon$ ring of Uranus (Elliot \& Nicholson 1984;
Goldreich \& Porco 1987; Porco \& Goldreich 1987; Chiang \& Goldreich 2000;
Mosqueira \& Estrada 2002)
as our showcase example. To what extent does the $\epsilon$ ring
satisfy the 3 requirements above?

\subsection{Ring Shepherding}
\label{rsmoon}
Ring confinement is manifestly occurring. The dissipation
length, $l_d$, for the $\epsilon$ ring whose optical depth is of order unity
is on the order of a scale height, $h$. Near the ring boundary,
we evaluate
$l_d \sim h \sim c_s/\Omega \sim (1 \cm/\ss)/(2\times 10^{-4} \rad/\ss)
\sim 50\m$ (see, e.g., Chiang \& Goldreich 2000, and references therein).
The ring spans a width $\Delta r \sim 60 \km$; thus $\Delta r / l_d \sim 10^3$.

\subsection{Slow Crossing}
\label{scmoon}
The $\epsilon$ ring is incredibly thin compared to its
protoplanetary disk counterpart; in the ring interior,
$h \sim 5\m$ (Chiang \& Goldreich 2000), while
$r \sim 5 \times 10^4 \km$.\footnote{Velocity dispersions
are $\sim$10$\times$ greater
near the ring boundary
than in the ring interior due to resonant perturbations by shepherd
satellites (Borderies, Goldreich, \& Tremaine 1982; Chiang \& Goldreich 2000).
In \S\ref{scmoon} and \S\ref{obssigna},
we evaluate $h$ in the ring interior because
the ring's intrinsic flux of angular momentum due to interparticle
collisions can be assessed there relatively undistorted
by torques exerted by satellites, while in \S\ref{rsmoon},
we take $h$ near the ring boundary because satellite-excited waves damp in that
vicinity.}
Moreover, the source of viscosity is relatively well understood
in rings (but see \S\ref{obssigna}); interparticle
collisions for a ring of optical depth unity provide
$\alpha \sim 1$ (Goldreich \& Tremaine 1978).
The $\epsilon$ ring shepherds, Cordelia and Ophelia,
have estimated masses of $M_1 \sim 5\times 10^{19}\g$ and $M_2 \sim 9\times
10^{19}\g$,
respectively, and orbital eccentricities of
$e_1 = 0.47 (\pm 0.41) \times 10^{-3}$, and
$e_2 = 10.1 (\pm 0.4) \times 10^{-3}$, respectively (Porco \& Goldreich 1987).
We assemble these data to re-normalize condition (\ref{reslow}) as

\begin{equation}
\label{slowmoon}
\max \, (\frac{M_1}{2\pi\Sigma r\Delta r},1) \gg 0.2 \, \frac{h/r}{10^{-7}} \,
\frac{h/\Delta r}{8\times 10^{-5}} \frac{10^{-3}}{e_1^{|q|}} \frac{\alpha}{1}
\frac{10^{-9}}{M_2/M_{U}} \, .
\end{equation}

\noindent
It is a remarkable coincidence
that the numerical coefficient on the right-hand side
of equation (\ref{slowmoon}) remains the same as that for
(\ref{reslow}) despite the vast difference in scales between
planetary rings and protoplanetary disks.
The mass of the $\epsilon$ ring is estimated to
be $2\times 10^{19}\g$ (Chiang \& Goldreich 2000).
Then the left-hand side of (\ref{slowmoon}) equals
$\max (3,1) = 3$. Thus, the requirement of
slow crossings is marginally satisfied.

\subsection{Fast Crossing}
\label{fcmoon}
We use the properties of the $\epsilon$ ring cited above
to re-normalize (\ref{refast}) as

\begin{equation}
\label{fastmoon}
2 \times 10^{-5} \left( \frac{\alpha}{1} \right)^{1/3} \left(
\frac{h/r}{10^{-7}} \right)^{2/3} \left( \frac{M_2/M_U}{10^{-9}} \right)^{1/2}
\left( \frac{M_1/M_U}{10^{-9}} \right)^{-2/3} \frac{e_1^{|q|/2}}{0.03} \ll 1 \,
{}.
\end{equation}

\noindent Thus, eccentricity changes directly
induced by the ring during resonance passage
can be ignored.

\subsection{Observational Signatures}
\label{obssigna}
Have resonance crossings left their imprint on Ophelia and Cordelia?
The present ratio of orbital semi-major axes of the two moons
is $(a_2/a_1)_0 = 1.08065 \pm 0.00004$ (Porco \& Goldreich 1987).
The nearest first-order resonance that lies
at a smaller ratio of semi-major axes is the
10:9 resonance: $(a_2/a_1)_{10:9} = 1.07277 \pm 10^{-6}$,
where the uncertainty measures the width
of the resonance as roughly estimated by equation (\ref{dares}).
To have crossed the 10:9 resonance, the satellite orbits must
have expanded by $\sim$$[(a_2/a_1)_0 - (a_2/a_1)_{10:9}]\,r \sim 400\km$.
Since this distance exceeds the radial width of the $\epsilon$ ring
($\Delta r \approx 60 \km$), which we take to be the maximum
length by which the ring-satellite system has expanded during its lifetime
(but see also the last paragraph of this section),
we conclude that
the moons could not have crossed any first-order resonance in the past.
This conclusion is consistent with Ophelia,
the more massive of the pair, currently possessing the greater
orbital eccentricity of the two; repeated resonance
crossings would predict the opposite to be true. In our view,
Ophelia's aberrantly large eccentricity is primordial.

What about future crossings?
The ratio of $(a_2/a_1)_0$ sits
closest to, but does not overlap with,
the 9:8 resonance: $(a_2/a_1)_{9:8} = 1.08169 \pm 10^{-6}$.
The satellite orbits must diverge by another
$\delta a_{9:8} \sim [(a_2/a_1)_{9:8} - (a_2/a_1)_0] \, r \sim 39 \km$
before resonance encounter.\footnote{The
9:8 resonance splits into two subresonances,
the so-called $e$ and $e'$ resonances (see Murray \& Dermott 1999).
The separation between them is 1 km and is due almost
entirely to the oblateness of Uranus. The widths of these
sub-resonances do not overlap.}
The event will be long in the waiting; the time of encounter lies
$\delta t_{9:8} = \delta a_{9:8} / \dot{a}_1 \sim 2 \times 10^4\yr$
in the future.\footnote{Tides raised by the
shepherds on the planet also engender divergent migration,
but over timescales of order $10^{10} \yr$.} Using relations derived
by Dermott, Malhotra, \& Murray (1988; see their Appendix B),
we estimate the perturbation eccentricity after resonance
crossing to be on the order of $6 \times 10^{-4}$ (= $e_{\rm crit}$
as defined by Dermott et al.~1988). This will be a minor
perturbation for Ophelia but will be relatively
significant for Cordelia.

What about the $\epsilon$ ring's direct effects on
satellite eccentricities?
Both first-order co-rotation torques and first-order Lindblad
torques operate simultaneously in the $\epsilon$ ring system,
resulting in eccentricity damping on
a timescale that is 22 $\times$ greater than that
given by equation (\ref{mineedot}) (GS;
Goldreich \& Tremaine 1980). We estimate
an exponential decay time for the satellite eccentricity
of $\sim$$3.6 \times 10^7\yr$, much longer
than either $\delta t_{9:8}$ or the likely age of the ring.\footnote{Tides
raised by the planet on the shepherds damp their orbital eccentricities
over $\sim$$8 \times 10^7\yr$.}

We note in passing that our expressions imply
that the $\epsilon$ ring is a young
creation of the solar system. For the ring
width to grow from $\Delta r/2$ to its
current width of $\Delta r$ would require
of order $\delta t_{\Delta r} \sim \delta t_{9:8} \Delta r / \delta a_{9:8}
\sim 3 \times 10^4\yr$.
If the viscous flux of angular momentum across
the ring midline were higher in the past,
we expect $\delta t_{\Delta r}$ to approximate well the age
of the ring. Even if the viscous flux were to remain constant
at all stages of ring evolution---as might be expected
if $\Sigma \nu \propto \Sigma / \tau$ were conserved,
where we have assumed that the ring's vertical optical
depth $\tau \gg 1$ in the ring's past---then we should multiply
$\delta t_{\Delta r}$ by a logarithm
that is unlikely to increase our estimate of the ring
age by more than an order of magnitude.
Goldreich \& Porco (1987) estimate an upper limit to the ring age
by dividing the present-day viscous torque into the angular momentum
required to be transferred from the inner shepherd to the outer shepherd
to separate their orbital semi-major axes by their current
difference of $\sim$4000 km. Using our estimate
for the viscous torque, we evaluate their upper limit
to be $\sim$$1\times 10^6\yr$. We consider this upper
limit to be a gross one, since it assumes the
satellite separation to be originally much smaller
than its current value.
In our view, a closer approximation to the ring age is obtained by
assuming the ring width to be originally much smaller than its current
value and the ring edges to occupy the same
first-order resonances with the shepherds as they do now,
in which case the initial satellite separation is within $\sim$98\% of
its current value. Additional uncertainty is introduced
by the magnitude of the ring viscosity. Our above
expressions employ the viscosity appropriate to a dilute
gas (Goldreich \& Tremaine 1978) of sound
speed $c_s \sim 0.1 \cm/\ss$ and turbulence
parameter $\alpha \approx \tau/(\tau^2 + 1) \approx 1$.
Then $\nu = \alpha c_s^2 / \Omega \approx 50 \cm^2/\ss$.
If ring particles are instead closely packed and behave
more like an incompressible liquid,
then the viscosity attains its minimum value
of $\Omega (\Sigma/\rho)^2 \approx 0.3 \cm^2/\ss$
(Borderies, Goldreich, \& Tremaine 1985),
where $\rho \approx 2 \gm/\cmm^{3}$ is
the internal density of a ring particle.
If this minimum viscosity obtains, then our estimate of the ring age,
$\sim \delta t_{\Delta r}$, and the
upper limit on the ring age obtained by
Goldreich \& Porco (1987), would need
to be revised upwards to $5\times 10^6\yr$
and $2\times 10^8\yr$, respectively. We close by
noting that (1) none of the four timescales we have computed,
ranging from $3\times 10^4 \yr$ to $2\times 10^8\yr$,
is as long as the age
of the solar system, $4\times 10^9\yr$,
and (2) none of our conclusions in \S\ref{scmoon}--\S\ref{fcmoon}
would change if the minimum viscosity were to obtain
and actual diffusion timescales were $\sim$$10^2$ times longer
than what we have assumed in those sections,
since requirement (\ref{slowmoon}) would be only
better satisfied and requirement
(\ref{fastmoon}) would remain satisfied.

\section{SUMMARY AND DISCUSSION}
\label{conc}

We have derived three requirements for the mutual excitation
of orbital eccentricities of two secondary
bodies by repeated resonance crossings
within a circumprimary disk,
a process originally proposed by Chiang, Fischer, \& Thommes (2002).

\begin{enumerate}
\item{For the orbits of two bodies to diverge, a ring of viscous material
must be shepherded between them. This requires that density waves excited
within the ring by each body damp near the edge where they are launched.}
\item{The divergent migration must be slow enough that the timescale
for traversal of the resonant width exceeds the resonant libration period.
In the context of Jovian-mass planets within a gaseous protoplanetary disk,
this requires that the ring mass be small compared to the planet mass
so that so-called ``Type II'' migration occurs over a timescale
longer than the viscous spreading time of the ring.}
\item{At the same time, the divergent migration must not be so slow
that the disk's direct effects on the bodies' eccentricities dominate
during resonance passage. This criterion appears well satisfied for
a variety of parameters because resonant widths
are typically small fractions of the orbital radius and the duration
of each passage is consequently short.}
\end{enumerate}

\noindent These necessary conditions are embodied in equations
(\ref{delrld}), (\ref{reslow}), and (\ref{refast}), respectively,
with normalizations appropriate to Jovian-mass planets in
protoplanetary disks.

The $\epsilon$ ring of Uranus and its attendant shepherds,
Cordelia and Ophelia, satisfy these requirements.
Unfortunately, these moons probably have yet to undergo
their first resonance crossing, so that we cannot
look to them for a signature of the process. In
$2 \times 10^4$--$3 \times 10^6\yr$,
we expect the radial diffusion of the $\epsilon$ ring
to wedge Cordelia and Ophelia through the 9:8 resonance,
after which Cordelia's eccentricity is likely to increase
by $\sim$$10^{-3}$.
Ophelia's eccentricity will probably
remain at its current high and likely primordial value
of $10^{-2}$. The range in timescales reflects
our uncertainty in the magnitude of the ring viscosity.
The Cassini spacecraft will probe
analogs of the $\epsilon$ ring around Saturn
(e.g., the Maxwell ringlet). We look forward to
sifting the abundance of new observations
of narrow ringlets around Saturn for signs of
divergent resonance crossings.

In the context of giant planets embedded within protoplanetary
disks, a shepherded ring of gas
in the vicinity of $r \sim 1 \AU$ and
having a surface density of $\Sigma \lesssim 500 \g/{\rm cm}^2$,
a dimensionless viscosity parameter of $\alpha \sim 0.1$,
and an aspect ratio of $h/r \approx 0.05$ can
drive two Jovian-mass planets through the
2:1 and higher-order resonances so that
their eccentricities magnify dramatically.
The above-cited upper limit to the surface density is a factor
of $\sim$3 lower than
that of the minimum-mass solar nebula; repeated
resonance crossings, in the context of the orbital
evolution of giant planets, is a process that is therefore restricted to the
later stages of the evolution of the protoplanetary disk
(ages $t \gtrsim 10^6\yr$).

Future work on this subject should incorporate numerical simulations
to verify that gaseous rings having the parameters that we have
outlined can indeed be shepherded by giant planets
and drive divergent migration. Perhaps our chief
uncertainty lies in the exact manner by which
density waves excited
by planets having masses $M > \max \, M_{\rm gap}$
dissipate. A potential outcome might be that
such waves, launched at one ring edge, release a great deal
of their angular momentum at the far ring edge,
rendering ring confinement impossible for
low-order resonant configurations such as the 2:1. It would
be worthwhile to measure the extent of eccentricity
excitation engendered by the crossings of only
high-order resonances, for which the requirements
of ring shepherding are less severe. Finally, divergent migration
within particle disks (see, e.g, Fernandez \& Ip 1984;
Murray et al.~1998) should also be explored, particularly
in light of our finding that divergent resonance crossings
are most effective when the ring mass is small
compared to the planet mass.

\acknowledgements
We thank Renu Malhotra for helpful discussions and pointing
us to previous work on divergent resonance crossings
in the context of the tidal migration of the
satellite pairs Miranda-Ariel and Enceladus-Tethys.
We also thank an anonymous referee for numerous
suggestions that significantly improved this paper.
This work was supported by National Science Foundation
Planetary Astronomy Grant AST-0205892, Hubble
Space Telescope Theory Grant HST-AR-09514.01-A,
and a Faculty Research Grant awarded by the University of California
at Berkeley.

\end{document}